\documentclass[preprint,showpacs,preprintnumbers,amsmath,amssymb,superscriptaddress]{revtex4}

\usepackage{graphicx,amsfonts}% Include figure files
\usepackage{epsfig}
\usepackage{dcolumn}% Align table columns on decimal point
\usepackage{bm}% bold math
\usepackage[english] {babel}

\begin{document}
\title{Electric charge quantization in $SU(3)_c \otimes SU(4)_L \otimes U(1)_X$ models.}
% \rule{0cm}{.6cm}}t

\author{J.M.\ Cabarcas}
\email{josecabarcas@usantotomas.edu.co}
\affiliation{Universidad Santo Tomas, Colombia}
\author{J.-Alexis Rodriguez}%
\email{jarodriguezl@unal.edu.co}
\affiliation{Departamento de F\'{\i}sica, Universidad Nacional de Colombia, Bogot\'a,
Colombia
}
\begin{abstract}
We obtain  electric charge quantization in the context of  models based on the gauge symmetry group $SU(3)_c \otimes SU(4)_L \otimes U(1)_X$.
The gauge models studied include three families to cancel out anomalies and a set of scalar fields to break 
spontaneously the  symmetry. To show the electric charge quantization, we use
clasical 
symmetry conditions and quantum quiral anomaly conditions. 
\end{abstract}


\maketitle

One of the most intriguing questions in the last decades is concerning  the electric charge quantization. It remains as an open question although there are some proposals.
The first one given by Dirac, which  includes magnetic monopoles  in a quantum mechanical theory  implying that electric charge is quantized \cite{dirac}. Another one
 came from grand unification theory using the group structure itself. It is based on  gauge models, that contain explicitly the $U(1)$ abelian group
in their structure and contributes to the $U(1)_{em}$ after the spontaneous symmetry breaking \cite{babu}. Using this last idea some studies have been done in the framework
of the $SU(3)_c \otimes SU(3)_L \otimes U(1)_X$ (331) models \cite{pires,long}. They use classical and quantum constraints to obtain the relationships between 
the $U(1)$ charges and  lead to electric charge quantization. It is relevant that in the standard model with one family the electric charge quantization can be obtained, but
when the three families are considered and with massless neutrinos then a dequantization arises up. It is possible to restore the electric charge quantization if Majorana neutrinos are
included.  This is related with the global hidden symmetry $U(1)_{B-L}$. On the other hand, in the framework of the 331 models the electric charge quantization is obtained
when  three families are involved all together and it does not depend on the neutrino  mass. Moreover if neutrinos are massive,
the charge electric quantization does not depend on the neutrino type, {\it i.e.,}  Dirac or Majorana type \cite{pires,long}.

%su4u1 motivation
The model 331 which enlarge the gauge group of the standard model share with the $SU(3)_c \otimes SU(4)_L \otimes U(1)_X$ (341) version the interesting feature of addressing
the problem of the number of fermion families \cite{pleitez,ponce2,sanchez}. It is concern to the anomaly cancellation among the families which is obtained when the number of left-handed triplets is equals
to the antitriplets in the 331 model \cite{331} and equal number of 4-plets and $4^*$-plets in the 341 model \cite{pleitez,ponce2,sanchez}. Taking into account the color degree of freedom. On the other hand,
if we add a right handed neutrino, one option is to have $\nu$, $e^-$, $\nu^c$ and $e^c$ in the same multiplet of $SU(4)_L$.  The gauge group $SU(4)$ is the highest symmetry group to be considered in
the electroweak sector as a consecuence of using the lightest leptons
as the particles which determine the gauge symmetry, each generator treated separately. Models based on the gauge symmetry $SU(3)_c \otimes SU(4)_L \otimes U(1)_X$ have
been studied before \cite{sanchez} and also this symmetry appears
in some Little Higgs models \cite{lh}. In this work, we study models based on $SU(3)_c \otimes SU(4)_L \otimes U(1)_X$ gauge group in order
to show how the electric charge quantization is satisfied.

%\section{Modelo 341}
We are going to consider models based on the gauge group $SU(3)_c \otimes SU(4)_L \otimes U(1)_X$.
The electric charge operator is defined as a linear combination of the diagonal generators of the group
\begin{eqnarray}
\label{charge}
Q= T_3+\beta T_8+\gamma T_{15}+X_f
\end{eqnarray}
where $T_i=\lambda_i/2 $ with $\lambda_i$ the Gell-Mann matrices for $SU(4)$ and $X_f$ the quantum number associated to $U(1)_X$. And the parameters $\beta$ and $\gamma$ define
the spectrum of the models, as we are going to show later on.
Usually in these models, in order to break spontaneously the gauge symmetry and 
give masses to the quark sector is necessary a set of four scalars
\begin{eqnarray}
 \chi\sim (1,4,X_\chi), \qquad \xi\sim (1,4,X_\xi), \qquad \eta\sim (1,4,X_\eta), \qquad {\rm and} \qquad \rho\sim (1,4,X_\rho)
\end{eqnarray}
with the following vacuum expectation values (VEV)
\begin{eqnarray}
 \chi= \begin{pmatrix}
        0 \cr 0 \cr 0\cr W/\sqrt{2}
       \end{pmatrix},
       \qquad 
 \xi=  \begin{pmatrix}
      0\cr 0 \cr V/\sqrt{2} \cr 0
       \end{pmatrix},
       \qquad
 \eta = \begin{pmatrix}
         0\cr w/\sqrt{2} \cr 0\cr
        \end{pmatrix},
        \qquad
 \rho= \begin{pmatrix}
        v/\sqrt{2}\cr 0\cr 0\cr 0
       \end{pmatrix} \, .
\end{eqnarray}
To verify that the electric
charge operator annihilates the VEVs  to have electric charge conservation,  we obtain
\begin{eqnarray}
\label{xscalars}
 X_\chi &=& \frac{3\gamma}{2 \sqrt{6}} \nonumber\\
 X_\xi  &=& \frac{\beta}{\sqrt{3}}-\frac{\gamma}{2\sqrt{6}} \nonumber\\
 X_\eta &=& \frac{1}{2}-\frac{\beta}{2\sqrt{3}}-\frac{\gamma}{2\sqrt{6}} \nonumber\\
 X_\rho &=& -\frac{1}{2}-\frac{\beta}{2\sqrt{3}}-\frac{\gamma}{2\sqrt{6}}
\end{eqnarray}
which also satisfies the relationship $X_\chi+X_\xi+X_\eta+X_\rho=0$.

On the other hand, the mass Lagrangian for the neutral gauge bosons $V^T=\begin{pmatrix} W^3_\mu & W^8_\mu & W^{15}_\mu & B_\mu \end{pmatrix}$ is
\begin{eqnarray}
{\cal L}_{mass}= \frac{1}{2}V^T\, M^2 \,V \, .
\end{eqnarray}
As it is usual, there is
a zero non-degenerate eigenvalue of the 
matrix  $M^2$ which is identified with the photon field $A_\mu$. In particular, the associated eigenvector is
\begin{eqnarray}
 A_\mu = \frac{g}{\sqrt{g^2+(1+ \beta^2+ \gamma^2)g^{'2}}}\left(\frac{g'}{g}W^3_\mu+
 \beta\frac{g}{g}W_\mu^8+\gamma\frac{g'}{g}W_\mu^{15}+B_\mu\right)
\label{foton}
 \end{eqnarray}
but we also have 
\begin{eqnarray}
 \begin{pmatrix}
  A_\mu & Z_\mu^1 & Z_\mu^2 & Z_\mu^3
 \end{pmatrix}^T = \begin{pmatrix}  
  W^3_\mu & W^8_\mu & W^{15}_\mu & B_\mu 
 \end{pmatrix} ^T U^T
\end{eqnarray}
where $U$ is $4 \times 4$ rotation matrix which can be written as
\begin{eqnarray}
 U= \begin{pmatrix}
     \frac{g'}{r} & U_{12} & U_{13} & U_{14}\cr
     \frac{\beta \,g'}{r} & U_{22} & U_{23} & U_{24}\cr
     \frac{\gamma g'}{r} & U_{32} & U_{33} & U_{34}\cr
     \frac{g}{r} & U_{42} & U_{43} & U_{44}
    \end{pmatrix}
\end{eqnarray}
with $r=\sqrt{g^2+(1+\beta^2+\gamma^2)g^{'2}} $. The elements $U_{ij}$ , $(i,j=2,3,4)$ are not relevant in this study.

Now, the equation (\ref{foton}) for the photon field implies that for an up quark type in the fundamental representation, the Lagrangian can be written as:
\begin{eqnarray}
 {\cal L}_{\bar u\, u\,\gamma}&&= {\bar{u_L}} \,i \, \gamma^\mu
 [\frac{ig}{2} U_{11} +\frac{i\,g}{2\sqrt{3}}U_{21} +\frac{i\,g}{2\sqrt{6}}U_{31}+ig'\,X_{4q} U_{41} ]A_\mu\ u_L
+\bar{u_R} \,i \, \gamma^\mu[ig'X_{u}^R\frac{g}{r} ]A_\mu\ u_R\nonumber\\
&& = -\frac{g\,g'}{r} \left[\frac{1}{2}+\frac{\beta}{2\sqrt{3}}+\frac{\gamma}{2\sqrt{6}}+X_{4q}\right]{\bar{u_L}}\, \gamma^\mu\,u_L A_\mu
 -\frac{g\,g'}{r} X_{u}^R \,{\bar{u_R}}\, \gamma^\mu\,u_R A_\mu
 \end{eqnarray}
but taking into account $X_\rho= -\frac{1}{2}-\frac{\beta}{2\sqrt{3}}-\frac{\gamma}{2\sqrt{6}}$, the Lagrangian takes the form
\begin{equation}
 {\cal L}_{\bar u \, u\,\gamma}= -\frac{g\,g'}{r}[X_{4q}-X_\rho]{\bar{u_L}}\, \gamma^\mu\,u_L 
A_\mu -\frac{g\,g'}{r}[X_{u}^R]{\bar{u_R}}\, \gamma^\mu\,u_R A_\mu
\end{equation}
and therefore asking for invariance under parity transformations, we arrive to $X_{u}^R=X_{4q}-X_\rho$.
Using the same arguments, we obtain for the heavy sector
\begin{eqnarray}
\label{15}
 X_{d}^R&=&X_{4q}-X_\eta \, ,\nonumber\\
 X_{J}^R&=&X_{4q}-X_\xi  \,  ,\nonumber\\
 X_{{\tilde{J}}}^R&=&X_{4q}-X_\chi  \, .
\end{eqnarray}

On the other hand, quarks can also be in the adjoint representation and for that case, we obtain
\begin{eqnarray}
X_{u'}^R&=&X_{4q*}+X_\eta \, , \nonumber\\ 
\label{19}
X_{d'}^R&=&X_{4q*}+X_\rho \, ,  \nonumber\\
 X_{J'}^R&=&X_{4q*}+X_\xi  \,  , \nonumber \\
 X_{{\tilde{J}'}}^R&=&X_{4q*}+X_\chi \, .
\end{eqnarray}

For the leptonic sector, the electromagnetic Lagrangian will be
\begin{eqnarray}
 {\cal L}_{\bar{\nu}\,\nu\,\gamma}&&= {\bar{\nu}_{L}} \,i \, \gamma^\mu
 [\frac{ig}{2}U_{11}+\frac{i\,g}{2\sqrt{3}}U_{21}+\frac{i\,g}{2\sqrt{6}}U_{31}+ig'\,X_{4\ell} U_{41} ]A_\mu\  \nu_{L}\nonumber\\
&&= -\frac{g\,g'}{r} 
\left[\frac{1}{2}+\frac{\beta}{2\sqrt{3}}+\frac{\gamma}{2\sqrt{6}}+X_{4\ell}\right]{\bar{\nu}_{L}}\, \gamma^\mu\,\nu_{L} 
A_\mu
 \end{eqnarray}
implying the relationship $X_{4\ell}=-\frac{1}{2}-\frac{\beta}{2\sqrt{3}}-\frac{\gamma}{2\sqrt{6}}=X_\rho$ under parity invariance.
Similarly for the other leptons including the heavy ones, we obtain
\begin{eqnarray}
 X_\ell^R&=&X_{4\ell}-X_\eta \, \, , \nonumber \\
  X_{F_\alpha}^R&=&X_{4\ell}-X_\xi \, \, , \nonumber \\
 X_{\tilde{F}_\alpha}^R&=&X_{4\ell}-X_\chi \,\, .
\end{eqnarray}

Up to now, we have dealt with classical symmetry conditions but we have also to consider the conditions coming from  the vanishing of the chiral anomaly coefficients. The relevant and
not trivial conditions in our particular case should satisfy
\begin{eqnarray}
 && \label{anomalia}\sum X_{\ell}^L+3\sum X_{q}^L=0 \, , \nonumber\\
 && 3\sum X_{q}^L-\sum_{sing}X_q^R =0 \, , \nonumber\\
 && 4\sum X_{\ell}^L+12\sum X_{q}^L-3\sum_{sing}X_q^R-\sum_{sing}X_\ell^R=0 \, , \\
 && 4\sum (X_{\ell}^L)^3+12\sum(X_{q}^L)^3-3\sum_{sing}(X_q^R)^3-\sum_{sing}(X_\ell^R)^3=0 \, \, . \nonumber
\end{eqnarray}
Considering explicitly the first equation (\ref{anomalia}), we obtain
\begin{eqnarray}
 3X_{4\ell}^L+3(X_{4q}^L-2X_{4q*}^L)=0
\end{eqnarray}
and using $X_{4\ell}^L=X_\rho$ then $X_{4q}^L=2X_{4q*}-X_\rho$. Additionally taking into account equations (\ref{15}) and (\ref{19}), we get
$
X_{4q}^L+X_{4q*}^L=X_\rho-X_\eta 
$
and therefore
\begin{eqnarray}
 X_{4q*}^L&=&\frac{2X_\rho-X_\eta}{3} \nonumber \\
 X_{4q}^L&=&\frac{X_\rho+2X_\eta}{3}  \, .
\end{eqnarray}
With a similar procedure, we obtain the following relations for the quark sector
\begin{eqnarray}
 X_u^R&=& \frac{2(X_\eta-X_\rho)}{3}  \, , \nonumber\\
 X_d^R&=& \frac{(X_\rho-X_\eta)}{3} \, , \nonumber\\
 X_J^R&=& \frac{X_\rho+2X_\eta-3X_\xi}{3} \, , \nonumber\\
 X_{\tilde{J}}^R&=& \frac{X_\rho+2X_\eta-3X_\chi}{3} \, \, .
\end{eqnarray}

These results are relevant for the electric charges because of the $X$ quantum numbers dependence of the scalar fields. Implying 
that the fermion electric charges are been quantized as a function of the scalar fields quantum numbers.

%\section{Caracterizacion de los modelos}

In the literature, many models based on the gauge symmetry 341 have been studied and the main differences are found on how the fermions
are assigned in the possible group representations \cite{pleitez,ponce2,sanchez}.  Different
values of the parameters $\beta$ and $\gamma$ in the charge operator (\ref{charge}) are fixed depending on how the fermions are in the multiplets of the group. 
The criterion to classify the possible models based on
341 symmetry is given by the values of the parameters $\beta$ and $\gamma$, generating models with or without exotic electric charges \cite{ponce2}. Now, we consider
the different associated fermion representations  for the 341 gauge symmetry models found in the literature. In general, fermions
are given by
\begin{eqnarray}
 q_{i}^T=\begin{pmatrix}
          d_i  & u_i & J_i & \tilde{J_i}
         \end{pmatrix}_L\sim (1,{\bf 4\ast}, X_{4q}) \nonumber
\end{eqnarray}
\begin{eqnarray}
 q_{3}^T=\begin{pmatrix}
          u_3  & d_3 & J_3 & \tilde{J_3}
         \end{pmatrix}_L\sim (1,{\bf 4}, X_{4q*}) \nonumber
\end{eqnarray}
\begin{eqnarray}
 l_{\alpha}^T=\begin{pmatrix}
          \nu_\alpha  & \ell_\alpha & F_\alpha & \tilde{F_\alpha}
         \end{pmatrix}_L\sim (1,{\bf 4}, X_{4\ell})
\end{eqnarray}
where the right singlets are not explicitly required because their quantization have been shown previoulsy through their $X_f^R$ numbers, equations (14) and (18).
Notice the relationship $Q(\ell)=Q(F_\alpha)$ which implies
\begin{eqnarray}
 -\frac{1}{2}+\frac{\beta}{2\sqrt{3}}+\frac{\gamma}{2\sqrt{6}}+X_{4\ell}
 =-\frac{\beta}{\sqrt{3}}+\frac{\gamma}{2\sqrt{6}}+X_{4\ell}
\end{eqnarray}
and therefore $\beta=\sqrt{3}/{3}$.
From the relationship $ Q(\ell)=Q(\tilde{F_\alpha})$, we have
\begin{eqnarray}
 -\frac{1}{2}+\frac{\beta}{2\sqrt{3}}+\frac{\gamma}{2\sqrt{6}}+X_{4\ell}
 &=&-\frac{3\gamma}{2\sqrt{6}}+X_{4\ell} 
\end{eqnarray}
and replacing the value  for $\beta=\sqrt{3}/3$, then we obtain $\gamma = \sqrt{6}/6$ .
The model concerning $ \beta=\sqrt{3}/{3}$ and $\gamma =\sqrt{6}/{6}$ has been studied by reference \cite{ponce2}
 and it corresponds to what they called model B.

Other type of models arise when electric charges satisfy $Q(\nu_\alpha)= Q(\tilde{F_\alpha})$, where
\begin{eqnarray}
 \frac{1}{2}+\frac{\beta}{2\sqrt{3}}+\frac{\gamma}{2\sqrt{6}}+X_{4\ell}
 =-\frac{3\gamma}{2\sqrt{6}}+X_{4\ell}
\end{eqnarray}
and taking $Q(\ell)=Q(F_\alpha)$ then we have

\begin{eqnarray}
 -\frac{1}{2}+\frac{\beta}{2\sqrt{3}}+\frac{\gamma}{2\sqrt{6}}+X_{4\ell}
 =-\frac{\beta}{\sqrt{3}}+\frac{\gamma}{2\sqrt{6}}+X_{4\ell}
\end{eqnarray}

and solving the equations, we obtain
\begin{eqnarray}
 \beta = \frac{\sqrt{3}}{3}\qquad {\rm and} \qquad \gamma=-\frac{2\sqrt{6}}{6} \, ,
\end{eqnarray}
this model  has been also studied previoulsy by Ponce and Sanchez \cite{ponce2} and it is named model F. As we can see, by fixing
the parameters $\beta$ and $\gamma$ and using them in (\ref {xscalars}) we obtain the $X$ numbers of the scalar sector and therefore the  $X_f$ numbers.

On the other hand, we have the model built up by Pisano and Pleitez \cite{pleitez} which choose the lepton sector in a different way
$
 l_{\alpha}^T=\begin{pmatrix}
          \nu_\alpha  & \ell_\alpha & \nu_\alpha^C & \ell_\alpha^C
         \end{pmatrix}_L
$
.  Following the analysis shown above, we 
notice that the electric charge operator for the leptonic sector can be written as $Q_l= T_3+\beta T_8+\gamma T_{15}+X_\rho$.
 Using equations (\ref{xscalars}),  the 
charge operator for the leptons arise naturally and it is
\begin{eqnarray}
 Q_l= \begin{pmatrix}
     0 & 0 & 0 & 0\cr
     0 & -1 & 0 & 0 \cr
     0 & 0 & -\frac{1}{2}-\frac{3\beta}{2\sqrt{3}}& 0\cr
     0 & 0 & 0 &-\frac{1}{2}-\frac{\beta}{2\sqrt{3}}-\frac{2\gamma}{\sqrt{6}}
    \end{pmatrix}
\end{eqnarray}
where it can be seen that the values for $\beta$ and $\gamma$ should be $ -1/\sqrt{3}$ and $-2\sqrt{6}/3$, respectively. 
And again electric charge is quantized. We should clarify that using the obtained values of $\beta$ and $\gamma$ in
equation (\ref{xscalars}) then $X_\xi=X_\rho=0$ and therefore they mix between them. Thus, the number of scalars is 
reduced to three and it is necessary to add a decuplet with $X=0$ number in order to give masses to all fermions of the spectrum \cite{pleitez}.

Models with and without exotic electric charges in the spectrum have been studied in the literature \cite{pleitez,islamabad,ponce2}. Reference \cite{ponce2} classifies the models without exotic
electric charges where they found eight different anomaly free models.  Models refered as models B and F have been analized here and we found 
that they quantized the electric charge. The same is true for models A and E which have a  spectrum in a representation completely congujate
 respect to models  B and F. These models A, B, E and F are three family models. On the other hand, on reference \cite{ponce2}, they also found five
  models (C, D, G,H and I) which
 cancel out the anomalies using only one or two families, but these are not realistic. Finally, regarding models that include exotic
 electric charges, we have studied here the case of a model by Pisano and Pleitez \cite{pleitez}. There is also another model presented in reference \cite{islamabad} 
 that includes right handed Majorana neutrinos in the spectrum but where the electric charge quantization is obtained analogously to the one 
 obtained in the model by Pisano and Pleitez \cite{pleitez}.

In summary, we have shown that the electric charge can be quantized in  models based on the gauge symmetry $SU(3)_c \otimes SU(4)_L \otimes U(1)_X$ that use
three families to cancel out the chiral anomalies. We have shown this quantization using electric charge conservation,
invariance under parity transformations and
chiral anomaly conditions. We have shown that electric charge quantization can be obtained in models based on 341 gauge symmetry, in models including exotic particles
 and also in models do not include exotic electric charges in the spectrum. 

We acknowledge V. Pleitez for the suggestions and discussion about this work and A. Leyva and C. Sandoval for the careful reading of the manuscript. J.A.R  
has been supported partially by grant 14844 DIB-UNAL Bogot\'a.

\end{document}